\documentclass[showpacs,twocolumn,preprintnumbers,amsmath,amssymb,superscriptaddress,nofootinbib]{revtex4-2}
\usepackage{graphicx}
\usepackage{bm}
\usepackage{ulem}
\usepackage{color}

\begin{document}

\newcommand{\rev}[1]{{\textcolor{blue}{#1}}}

\newcommand{\simgt}{\lower.5ex\hbox{$\; \buildrel > \over \sim \;$}}
\newcommand{\simlt}{\lower.5ex\hbox{$\; \buildrel < \over \sim \;$}}
\newcommand{\sang}[1]{\left\langle#1\right\rangle}

\title{Amplitude and phase fluctuations of gravitational waves magnified by strong gravitational lensing}

\author{Masamune Oguri}
\affiliation{
Center for Frontier Science, Chiba University, 1-33 Yayoi-cho, Inage-ku, Chiba 263-8522, Japan}
\affiliation{
Department of Physics, Graduate School of Science, Chiba University, 1-33 Yayoi-Cho, Inage-Ku, Chiba 263-8522, Japan
}
\author{Ryuichi Takahashi}
\affiliation{
Faculty of Science and Technology, Hirosaki University, 3 Bunkyo-cho, Hirosaki, Aomori 036-8561, Japan
}

\date{\today}

\begin{abstract}
We discuss how small-scale density perturbations on the Fresnel scale
affect amplitudes and phases of gravitational waves that are
magnified by gravitational lensing in geometric optics. We derive
equations that connect the small-scale density perturbations with the
amplitude and phase fluctuations to show that such perturbative wave
optics effects are significantly boosted in the presence of macro model
magnifications such that amplitude and phase fluctuations can easily
be observed for highly magnified gravitational waves. We discuss
expected signals due to microlensing by stars, dark matter
substructure, fuzzy dark matter, and primordial black holes. 
\end{abstract}


\maketitle

\section{Introduction}

The first direct observation of gravitational waves from a compact
binary merger in 2015 opened a new window to study the Universe
\cite{LIGOScientific:2016aoc}. One of the unique characteristics of
gravitational waves from compact binary mergers is that their
wavelengths are typically much longer than those of any electromagnetic
observations. The longer wavelengths indicate that wave optics
effects in the propagation of gravitational waves can become more
important, unlike electromagnetic observations for which the geometric
optics approximation can safely be used in most cases. Indeed, it has
been shown that wave optics effects play a key role in predicting
gravitational lensing effects on gravitational waves in several cases
(see e.g., \cite{Nakamura:1999,Oguri:2019fix} for reviews). 

When wave optics effects are relevant, the propagation of
gravitational waves is barely affected by any structures smaller than the
Fresnel scale due to the diffraction. Since the Fresnel scale
\cite{Macquart:2004sh,Takahashi:2005ug}
depends on the frequency of gravitational waves as $\propto f^{-1/2}$, the
amplification factor due to gravitational lensing depends also on the
frequency in wave optics. Therefore, by checking the frequency
evolution of gravitational waveforms we can in principle identify the
signature of wave optics gravitational lensing for an ensemble of
events or even for a single event (e.g., \cite{Nakamura:1997sw}).   

With this situation in mind, Takahashi \cite{Takahashi:2005ug}
proposed a novel method to probe small-scale density fluctuations from
amplitude and phase fluctuations of gravitational waves. Adopting the
Born approximation \cite{Takahashi:2005sxa}, simple relations between
the matter power spectrum and amplitude and phase fluctuations are
derived. Oguri and Takahashi \cite{Oguri:2020ldf} extended this work
and explored the possibility of using amplitude and phase fluctuations
as a probe of dark low-mass halos and primordial black holes. Inamori
and Suyama \cite{Inamori:2021tlx} reported a simple universal relation
between amplitude and phase fluctuations.

In this paper, we ask how wave optics effects modify amplitudes and
phases of gravitational waves magnified by strong gravitational lensing 
due to normal galaxies or clusters of galaxies. Observing such strong
lensing events is expected to be within a reach of ongoing ground
experiments
(e.g., \cite{Ng:2017yiu,Broadhurst:2018saj,Li:2018prc,Oguri:2018muv,Cusin:2019eyv,Mukherjee:2020tvr,LIGOScientific:2021izm,Diego:2021fyd,Yang:2021viz,Xu:2021bfn,Mukherjee:2021qam,Wierda:2021upe}),
for which the geometric optics approximation can be
safely used. Wave optics effects on multiple images due to
microlensing by stars have been studies in detail
\cite{Diego:2019lcd,Cheung:2020okf,Mishra:2021xzz,Yeung:2021roe}.
Here we derive simple analytic expressions that connect amplitude
and phase fluctuations and small-scale density fluctuations on the
Fresnel scale in such situations. We show how the macro model
magnification enhances the amplitude and phase fluctuations such that
they can be easily detected when the images are highly magnified.

This paper is organized as follows.
In Sec.~\ref{sec:derivation} we derive expressions of amplitude and
phase fluctuations in the presence of the macro model magnification.
In Sec.~\ref{sec:discussions} we discuss amplitude and phase
fluctuations computed by our formalism, including calculations of
expected signals in several models. We summarize our results in
Sec.~\ref{sec:summary}.
Throughout the paper we assume a flat Universe with the matter density
$\Omega_M=0.3$, the cosmological constant $\Omega_\Lambda=0.7$, and
the dimensionless Hubble constant $h=0.7$.

\section{Derivation}\label{sec:derivation}

\subsection{Complex amplification factor in the presence of small-scale perturbations}

We define a complex amplification factor $F$ by the ratio of waveforms
with and without gravitational lensing for monochromatic waves 
$\psi(\boldsymbol{x},t)=\tilde{\psi}(\boldsymbol{x})e^{-2\pi if t}$
with frequency $f$. Specifically, 
\begin{equation}
F=\frac{\tilde{\psi}^{\rm L}}{\tilde{\psi}},
\end{equation}
where $\tilde{\psi}$ denotes the original waveform without
gravitational lensing and $\tilde{\psi}^{\rm L}$ is the observed amplitude
including gravitational lensing effects. For simplicity, throughout
the paper we consider gravitational lensing effects in a single thin
lens plane at redshift $z$. Under the flat sky approximation, the
complex amplification factor is described by
(e.g., \cite{Nakamura:1999,Oguri:2019fix})
\begin{equation}
  F_0(f, \boldsymbol{q}_\beta)
  =\frac{\chi_{\rm s}}{c\chi(\chi_{\rm s}-\chi)}\frac{f}{i}
  \int d\boldsymbol{q}\,e^{2\pi i f \Delta
    t_0(\boldsymbol{q}, \boldsymbol{q}_\beta)},
\end{equation}
\begin{equation}
  \Delta t_0(\boldsymbol{q}, \boldsymbol{q}_\beta)
  =\frac{\chi_{\rm s}}{c\chi(\chi_{\rm s}-\chi)}
  \left[\frac{(\boldsymbol{q}-\boldsymbol{q}_\beta)^2}{2}-\phi_0(\boldsymbol{q})\right],
\end{equation}
where $\chi$ and $\chi_{\rm s}$ are comoving radial distances to the
lens and the source, respectively, $\boldsymbol{q}$ denotes the 
comoving two-dimensional coordinates in the lens plane, 
$\boldsymbol{q}_\beta$ is the source position on the sky projected on
the lens plane, and $\phi_0(\boldsymbol{q})$ is the lens potential.

We consider a small region around a $j$-th multiple image at
$\boldsymbol{q}_j$ for a strong lens system. In the geometric optics
limit ($|f|\rightarrow \infty$), the complex amplification factor for
this image is derived as 
\begin{equation}
  F_{\rm G0}^j(f, \boldsymbol{q}_\beta)
  \simeq |\mu_0(\boldsymbol{q}_j)|^{1/2}
  e^{2\pi i f \Delta t_0(\boldsymbol{q}_j, \boldsymbol{q}_\beta)}
  e^{-i\pi n_j{\rm sgn}(f)},
\end{equation}
where $\mu_0(\boldsymbol{q}_j)$ is the signed magnification factor and
$n_j=0$, $1/2$, $1$ for minimum, saddle point, and maximum images. The
factor ${\rm sgn}(f)$ ensures that lensing of a real wave packet
remains real \cite{Ezquiaga:2020gdt}.
  
Now we consider small perturbations on the main lens potential 
\begin{equation}
  \phi(\boldsymbol{q})=\phi_0(\boldsymbol{q})+\phi'(\boldsymbol{q}),
\end{equation}
\begin{equation}
  F(f, \boldsymbol{q}_\beta)
  =\frac{\chi_{\rm s}}{c\chi(\chi_{\rm s}-\chi)}\frac{f}{i}
  \int d\boldsymbol{q}\,e^{2\pi i f \Delta
    t(\boldsymbol{q}, \boldsymbol{q}_\beta)},
\end{equation}
where the time delay surface can be decomposed into
$\Delta t=\Delta t_0+\Delta t'$ with
\begin{equation}
  \Delta t'(\boldsymbol{q})
  =-\frac{\chi_{\rm s}}{c\chi(\chi_{\rm s}-\chi)}\phi'(\boldsymbol{q}).
\end{equation}
By adding $\phi'(\boldsymbol{q})$ to the lens potential, the $j$-th
image position is in general shifted from $\boldsymbol{q}_j$ to
$\boldsymbol{q}_j+\boldsymbol{r}_j$. The complex amplification factor
in the geometric optics limit is then modified as
\begin{equation}
  F_{\rm G}^j(f, \boldsymbol{q}_\beta)
  \simeq |\mu(\boldsymbol{q}_j+\boldsymbol{r}_j)|^{1/2}
  e^{2\pi i f \Delta t(\boldsymbol{q}_j+\boldsymbol{r}_j, \boldsymbol{q}_\beta)}
  e^{-i\pi n_j{\rm sgn}(f)},
\end{equation}
Since $\Delta t(\boldsymbol{q}_j+\boldsymbol{r}_j, \boldsymbol{q}_\beta)
\simeq \Delta t(\boldsymbol{q}_j,\boldsymbol{q}_\beta)$ to the first
order approximation, the phase shift $\delta_j$ due to the perturbation
in the geometric optics limit is given by 
\begin{equation}
  \delta_j={\rm arg}\left(\frac{F_{\rm G}^j}{F_{\rm G0}^j}\right)\simeq
  -2\pi f\frac{\chi_{\rm s}}{c\chi(\chi_{\rm s}-\chi)}\phi'(\boldsymbol{q}_j).
\end{equation}

We consider the situation that the time delay perturbation $\Delta
t'$ is sufficiently small, $|f\Delta t'|\ll 1$, while the geometric
optics approximation can be applied to $\Delta t_0$, i.e., we
include wave optics effects perturbatively. In this case, the complex
amplification factor for the $j$-th image is modified as 
\begin{align}
  e^{-i\delta_j}F^j(f, \boldsymbol{q}_\beta)
  =&\frac{\chi_{\rm s}}{c\chi(\chi_{\rm s}-\chi)}\frac{f}{i}
  \int d\boldsymbol{q}\nonumber\\
  &\times e^{2\pi i f \left\{\Delta
    t_0(\boldsymbol{q}, \boldsymbol{q}_\beta)+\Delta
    t'(\boldsymbol{q})
    -\Delta t'(\boldsymbol{q}_j)\right\}}\nonumber\\
  \simeq&F_{\rm G0}^j(f, \boldsymbol{q}_\beta)
  -2\pi f^2\left\{\frac{\chi_{\rm s}}{c\chi(\chi_{\rm s}-\chi)}\right\}^2 
  \nonumber\\
  &\times \int d\boldsymbol{q}\,
  \left\{\phi'(\boldsymbol{q})-\phi'(\boldsymbol{q}_j)\right\}\,
  e^{2\pi i f \Delta t_0(\boldsymbol{q}, \boldsymbol{q}_\beta)}.
  \label{eq:f_tot}
\end{align}
We then consider the local coordinates $\boldsymbol{r}$ around
$\boldsymbol{q}_j$ such that $\boldsymbol{q}=\boldsymbol{q}_j+\boldsymbol{r}$.
Since $\boldsymbol{q}_j$ is a stationary point of the time delay
surface $\Delta t_0(\boldsymbol{q}, \boldsymbol{q}_\beta)$, it satisfies
\begin{equation}
  \left.\frac{\partial\Delta t_0}{\partial \boldsymbol{q}}
  \right|_{\boldsymbol{q}=\boldsymbol{q}_j}=0.
\end{equation}
Hence we can expand $\Delta t_0$ around $\boldsymbol{q}_j$ as
\begin{equation}
  \Delta t_0(\boldsymbol{q},
  \boldsymbol{q}_\beta) \simeq  \Delta t_0(\boldsymbol{q}_j,
  \boldsymbol{q}_\beta)+\frac{1}{2}\boldsymbol{r}^\top\boldsymbol{{\rm
      H}}(\boldsymbol{q}_j)\boldsymbol{r}+\cdots,
  \label{eq:delta_t0}
\end{equation}
\begin{equation}
  H_{ab}(\boldsymbol{q}_j)=\left.\frac{\partial^2 \Delta t_0}{\partial
      q_a\partial q_b}\right|_{\boldsymbol{q}=\boldsymbol{q}_j}.
\end{equation}
Since the Hessian matrix $\boldsymbol{{\rm H}}$ is symmetric it can be
diagonazied by rotating the coordinate system and is described as
\begin{equation}
  H_{ab}(\boldsymbol{q}_j)=
\frac{\chi_{\rm s}}{c\chi(\chi_{\rm s}-\chi)}
\begin{pmatrix}
  \mu_{j,1}^{-1} & 0 \\
  0 & \mu_{j,2}^{-1}  
\end{pmatrix},
\end{equation}
where $\mu_{j,1}$ and $\mu_{j,2}$ satisfy
$\mu_{j,1}\mu_{j,2}=\mu_0(\boldsymbol{q}_j)$. We refer to
$\mu_0(\boldsymbol{q}_j)$ as the macro model magnification
throughout the paper. By inserting this expression of the Hessian
matrix $\boldsymbol{{\rm H}}$, Eq.~(\ref{eq:delta_t0}) reduces to
\begin{align}
  \Delta t_0(\boldsymbol{q},
  \boldsymbol{q}_\beta) \simeq  &\Delta t_0(\boldsymbol{q}_j,
  \boldsymbol{q}_\beta)\nonumber\\
   &+\frac{\chi_{\rm s}}{2c\chi(\chi_{\rm s}-\chi)}
  \left(\mu_{j,1}^{-1}r_1^2+\mu_{j,2}^{-1}r_2^2\right)
  +\cdots,
\end{align}
where $\boldsymbol{r}=(r_1,\,r_2)$. Eq.~(\ref{eq:f_tot}) then reduces to
\begin{align}
  e^{-i\delta_j}F^j(f, \boldsymbol{q}_\beta)
  \simeq &
  F_{\rm G0}^j(f, \boldsymbol{q}_\beta)
  -\frac{1}{2\pi r_{\rm F}^4}
  e^{2\pi i f \Delta t_0(\boldsymbol{q}_j, \boldsymbol{q}_\beta)}
  \nonumber\\
  &\times \int d\boldsymbol{r}\,
 \left\{\phi'(\boldsymbol{r})-\phi'(\boldsymbol{0})\right\}
 \nonumber\\
 &\times \exp\left(i \frac{\mu_{j,1}^{-1}r_1^2+\mu_{j,2}^{-1}r_2^2}{2 r_{\rm
      F}^2}\right),
  \label{eq:f_tot_2}
\end{align}
where the Fresnel scale is introduced as
\begin{equation}
r_{\rm F}=\sqrt{\frac{c\chi(\chi_{\rm s}-\chi)}{2\pi f\chi_{\rm s}}}.
\label{eq:r_fresnel}
\end{equation}

In evaluating Eq.~(\ref{eq:f_tot_2}), it is useful to consider the
Fourier transform of $\phi'(\boldsymbol{r})$ 
\begin{equation}
\phi'(\boldsymbol{r})=\int \frac{d\boldsymbol{k}}{(2\pi)^2}\,
\tilde{\phi}'(\boldsymbol{k})
e^{i\boldsymbol{k}\cdot\boldsymbol{r}}.
\end{equation}
In the Fourier space, $\tilde{\phi}'(\boldsymbol{k})$ and 
convergence $\tilde{\kappa}_j(\boldsymbol{k})$ are related with each other by
\begin{equation}
  -k^2\tilde{\phi}'(\boldsymbol{k})=2\tilde{\kappa}_j(\boldsymbol{k}),
\end{equation}
where subscript $j$ is added to convergence to make it clear that it is
the small-scale convergence field in the vicinity of
$\boldsymbol{q}_j$, which is defined to have zero mean i.e.,
$\langle \kappa\rangle=0$. Inserting these expressions, we obtain
\begin{align}
e^{-i\delta_j}F^j(f, \boldsymbol{q}_\beta)
  &\simeq
F_{\rm G0}^j(f, \boldsymbol{q}_\beta)\nonumber\\
&\;\;\;+|\mu_{j,1}\mu_{j,2}|^{1/2}e^{-i\pi n_j{\rm sgn}(f)}
e^{2\pi i f \Delta t_0(\boldsymbol{q}_j, \boldsymbol{q}_\beta)}
\nonumber\\
&\times \int \frac{d\boldsymbol{k}}{(2\pi)^2}\,
  \tilde{\kappa}_j(\boldsymbol{k})
  \frac{i}{r_{\rm F}^2k^2/2}
  \nonumber\\
&\times \left[\exp\left(-i\frac{\mu_{j,1}r_{\rm F}^2}{2}k_1^2-i\frac{\mu_{j,2}r_{\rm
      F}^2}{2}k_2^2\right)-1\right]\nonumber\\
&=F_{\rm G0}^j(f, \boldsymbol{q}_\beta)\left[1+\int \frac{d\boldsymbol{k}}{(2\pi)^2}\,
  \tilde{\kappa}_j(\boldsymbol{k})\tilde{G}_j(\boldsymbol{k}, f)\right],
\end{align}
where
\begin{equation}
  \tilde{G}_j(\boldsymbol{k}, f)= \frac{i}{r_{\rm F}^2k^2/2}
\left[\exp\left(-i\frac{\mu_{j,1}r_{\rm F}^2}{2}k_1^2-i\frac{\mu_{j,2}r_{\rm
      F}^2}{2}k_2^2\right)-1\right].
\end{equation}

From the explicit expressions of the complex amplification factors, we
also obtain
\begin{equation}
\frac{F_{\rm G}^j(f, \boldsymbol{q}_\beta)}{F_{\rm G0}^j(f, \boldsymbol{q}_\beta)}
=\frac{|\mu(\boldsymbol{q}_j+\boldsymbol{r}_j)|^{1/2}}{|\mu_0(\boldsymbol{q}_j)|^{1/2}}
e^{i\delta_j}.
\end{equation}
By using this expression, we finally obtain
\begin{align}
  F^j(f, \boldsymbol{q}_\beta)
  &\simeq\frac{|\mu_0(\boldsymbol{q}_j)|^{1/2}}{|\mu(\boldsymbol{q}_j+\boldsymbol{r}_j)|^{1/2}}
  F_{\rm G}^j(f, \boldsymbol{q}_\beta)
  \nonumber\\
 & \times \left[1+\int \frac{d\boldsymbol{k}}{(2\pi)^2}\,
  \tilde{\kappa}_j(\boldsymbol{k})\tilde{G}_j(\boldsymbol{k},
  f)\right]\nonumber\\
&=|\mu_0(\boldsymbol{q}_j)|^{1/2}
e^{2\pi i f \Delta t(\boldsymbol{q}_j+\boldsymbol{r}_j, \boldsymbol{q}_\beta)}
\nonumber\\
&\times e^{-i\pi n_j{\rm sgn}(f)}\left[1+\int \frac{d\boldsymbol{k}}{(2\pi)^2}\,
  \tilde{\kappa}_j(\boldsymbol{k})\tilde{G}_j(\boldsymbol{k},
  f)\right].
\end{align}
This is a general expression of the complex amplification factor in
the presence of small-scale perturbations on the Fresnel scale. We note
that this expression can be applied to not only multiple images for a
strongly lensed system but also single image systems. We also note
that previous work \cite{Takahashi:2005ug,Oguri:2020ldf,Inamori:2021tlx}
essentially corresponds to a special situation with
$\mu_{j,1}=\mu_{j,2}=1$. 

It is worth noting that $F^j$ is not a direct observable, because an
intrinsic, unlensed waveform is usually unknown. However, waveforms of
compact binary mergers are parameterized by a small number of physical
parameters of binaries including mass and spin as well as the
configuration of the detector with respect to the direction of the
source, and any deviations from physical templates may be ascribed to
wave optics effects in the propagation of gravitational waves.
Previous work \cite{Dai:2018enj,Choi:2021bkx} explored the possibility
of using such wave optics signature in individual binary merger
waveforms to probe small-mass subhalos.   

\subsection{Amplitude and phase fluctuations for multiply imaged
  gravitational waves}

Here we discuss an alternative approach to detect amplitude and phase
fluctuations by comparing waveforms of multiple images. In this case,
by comparing waveform shapes of $l$-th and $m$-th multiple images
with their time delay $\Delta t_{lm}$, which should be determined from
the data, we can measure the ratio of complex magnification factors
that is is independent of an intrinsic waveform. Specifically,
we define the ratio as
\begin{align}
R_{lm}(f)&=e^{-2\pi if\Delta t_{lm}}
\frac{F^l(f,\boldsymbol{q}_\beta)}{F^m(f,\boldsymbol{q}_\beta)}\nonumber\\
&=\frac{|\mu_0(\boldsymbol{q}_l)|^{1/2}}{|\mu_0(\boldsymbol{q}_m)|^{1/2}}
e^{-i\pi (n_l-n_m){\rm sgn}(f)}\left[1+\eta_{lm}(f)\right],
\end{align}
where
\begin{equation}
\eta_{lm}(f)=\int \frac{d\boldsymbol{k}}{(2\pi)^2}\,
  \left[\tilde{\kappa}_l(\boldsymbol{k})\tilde{G}_l(\boldsymbol{k},f)
  -\tilde{\kappa}_m(\boldsymbol{k})\tilde{G}_m(\boldsymbol{k},f)\right].
\end{equation}
The function $\eta_{lm}(f)$ describes effects of small-scale
perturbations, which for instance induces additional phase
shift on top of the phase shift due to the Morse index
\cite{Ezquiaga:2020gdt,Dai:2017huk,Dai:2020tpj}, and represents
perturbative wave optics effects. We further decompose
$\eta_{lm}(f)$ into amplitude $K_{lm}(f)$ and phase fluctuations
$S_{lm}(f)$ as
\begin{equation}
1+\eta_{lm}(f)\simeq \left[1+K_{lm}(f)\right]e^{iS_{lm}(f)}.
\end{equation}
We note that small-scale perturbations on $l$-th and $m$-th
multiple images, which represent local density fluctuations with zero
mean around each multiple image, are to a good approximation regarded
as statistically independent because the transverse separation between
these multiple images (typically $\gtrsim$~kpc for galaxy-scale strong
lensing) is usually much larger than the Fresnel scale
($\sim$~pc for $f\sim 10$~Hz).
The convergence power spectra $P_\kappa(k)$ can be defined as  
\begin{equation}
\langle \tilde{\kappa}_l(\boldsymbol{k})\tilde{\kappa}_m(\boldsymbol{k}')\rangle
=\delta_{lm}(2\pi)^2\delta^{\rm D}(\boldsymbol{k}+\boldsymbol{k}')P^l_\kappa(k),
\end{equation}
where $\delta_{lm}$ denotes the Kronecker delta.
Using this relation, we can compute dispersions of $K_{lm}(f)$ and $S_{lm}(f)$ as
\begin{equation}
 \langle K_{lm}^2(f)\rangle = \langle K_{l}^2(f)\rangle +\langle K_{m}^2(f)\rangle,
\end{equation}
\begin{align}
 \langle K_{j}^2(f)\rangle = &\int \frac{d\boldsymbol{k}}{(2\pi)^2}\,P^j_\kappa(k)
 \nonumber\\
 &\times\left[\frac{\sin(\mu_{j,1}r_{\rm F}^2k_1^2/2+\mu_{j,2}r_{\rm
      F}^2k_2^2/2)}{r_{\rm F}^2k^2/2}\right]^2,
\label{eq:kj1}
\end{align}
\begin{equation}
 \langle S_{lm}^2(f)\rangle = \langle S_{l}^2(f)\rangle +\langle S_{m}^2(f)\rangle,
\end{equation}
\begin{align}
 \langle S_{j}^2(f)\rangle = &\int \frac{d\boldsymbol{k}}{(2\pi)^2}\,P^j_\kappa(k)
 \nonumber\\
 &\times\left[\frac{\cos(\mu_{j,1}r_{\rm F}^2k_1^2/2+\mu_{j,2}r_{\rm
      F}^2k_2^2/2)-1}{r_{\rm F}^2k^2/2}\right]^2,
\label{eq:sj1}
\end{align}
assuming the statistical independence for the $l$-th and $m$-th
multiple images as discussed above.
We find that the integrations over the polar angle in the
$\boldsymbol{k}$-space can be performed using the Bessel
function. Specifically, by introducing the following quantities 
\begin{equation}
 A=r_{\rm F}^2k^2/2,
\end{equation}
\begin{equation}
 A^{j,\pm}=(\mu_{j,1}\pm\mu_{j,2})r_{\rm F}^2k^2/2,
\end{equation}
to simplify the expressions, Eqs.~(\ref{eq:kj1}) and (\ref{eq:sj1})
reduce to 
\begin{equation}
 \langle K_{j}^2(f)\rangle = \int \frac{k\,dk}{2\pi}\,P^j_\kappa(k)
 I_K^j,
\label{eq:k2_j}
\end{equation}
\begin{equation}
 I_K^j=\frac{1-J_0(A^{j,-})\cos(A^{j,+})}{2A^2},
\label{eq:i_k}
\end{equation}
\begin{equation}
 \langle S_{j}^2(f)\rangle = \int \frac{k\,dk}{2\pi}\,P^j_\kappa(k)
 I_S^j,
\label{eq:s2_j}
\end{equation}
\begin{equation}
 I_S^j=
 \frac{3-4J_0(A^{j,-}/2)\cos(A^{j,+}/2)+J_0(A^{j,-})\cos(A^{j,+})}{2A^2}.
\label{eq:i_s}
\end{equation}
The amplitude shift $K_{lm}(f)$ is not observable from observations in
a single frequency $f$ because it degenerates with the macro model
magnification $\mu_0$. On the other hand, even for observations in a
single frequency, the phase shift $S_{lm}(f)$ can in principle
directly be observed in a manner similar to the measurement of the
Morse phase (see e.g., \cite{Dai:2020tpj}). 

\subsection{Frequency evolution of amplitude and phase fluctuations}

In previous work \cite{Takahashi:2005ug,Oguri:2020ldf}, the frequency
evolution of amplitude and phase fluctuations are considered as a
means of measuring the wave optics effects. Here we consider the
evolution of $K_{lm}(f)$ and $S_{lm}(f)$  between frequencies $f_1$
and $f_2$. Again, by assuming the statistical independence between
small-scale fluctuations around $l$-th and $m$-th images, we can
compute their dispersions as
\begin{align}
 \langle \left[K_{lm}(f_1)-K_{lm}(f_2)\right]^2\rangle = &\langle
 \left[K_{l}(f_1)-K_{l}(f_2)\right]^2\rangle \nonumber\\
 &+\langle \left[K_{m}(f_1)-K_{m}(f_2)\right]^2\rangle,
\end{align}
\begin{align}
 \langle \left[S_{lm}(f_1)-S_{lm}(f_2)\right]^2\rangle = &\langle
 \left[S_{l}(f_1)-S_{l}(f_2)\right]^2\rangle \nonumber\\
 &+\langle \left[S_{m}(f_1)-S_{m}(f_2)\right]^2\rangle.
\end{align}
Writing
\begin{equation}
 A_1=\left\{r_{\rm F}(f_1)\right\}^2k^2/2,
\end{equation}
\begin{equation}
 A^{j,\pm}_1=(\mu_{j,1}\pm\mu_{j,2})\left\{r_{\rm F}(f_1)\right\}^2k^2/2,
\end{equation}
and similar quantities for $f_2$ as $A_2$ and $A_2^{j,\pm}$, 
we find that the integrations over the polar angle in the
$\boldsymbol{k}$-space can be performed to obtain
\begin{equation}
 \langle \left[K_{j}(f_1)-K_{j}(f_2)\right]^2\rangle =
  \int \frac{k\,dk}{2\pi}\,P^j_\kappa(k)I_K^{j,12},
\end{equation}
\begin{equation}
 \langle \left[S_{j}(f_1)-S_{j}(f_2)\right]^2\rangle =
  \int \frac{k\,dk}{2\pi}\,P^j_\kappa(k)I_S^{j,12},
\end{equation}
where
\begin{widetext}
\begin{align}
  I_K^{j,12}=&\frac{1-J_0(A_1^{j,-})\cos(A_1^{j,+})}{2A_1^2}
  +\frac{1-J_0(A_2^{j,-})\cos(A_2^{j,+})}{2A_2^2}+\frac{J_0((A_1^{j,-}+A_2^{j,-})/2)\cos((A_1^{j,+}+A_2^{j,+})/2)}{A_1A_2}\nonumber\\
  &-\frac{J_0((A_1^{j,-}-A_2^{j,-})/2)\cos((A_1^{j,+}-A_2^{j,+})/2)}{A_1A_2},
\end{align}
\begin{align}
  I_S^{j,12}=&\frac{3-4J_0(A_1^{j,-}/2)\cos(A_1^{j,+}/2)+J_0(A_1^{j,-})\cos(A_1^{j,+})}{2A_1^2}
    +\frac{3-4J_0(A_2^{j,-}/2)\cos(A_2^{j,+}/2)+J_0(A_2^{j,-})\cos(A_2^{j,+})}{2A_2^2}
    \nonumber\\
    &
    -\frac{2}{A_1A_2}
    -\frac{J_0((A_1^{j,-}+A_2^{j,-})/2)\cos((A_1^{j,+}+A_2^{j,+})/2)}{A_1A_2}
    -\frac{J_0((A_1^{j,-}-A_2^{j,-})/2)\cos((A_1^{j,+}-A_2^{j,+})/2)}{A_1A_2}\nonumber\\
   &
    +\frac{2J_0(A_1^{j,-}/2)\cos(A_1^{j,+}/2)}{A_1A_2}
    +\frac{2J_0(A_2^{j,-}/2)\cos(A_2^{j,+}/2)}{A_1A_2}.
\end{align}
\end{widetext}
To conclude, we find simple analytic expressions for dispersions of
amplitude and phase shifts due to small-scale perturbations on the
Fresnel scale.

\begin{figure}[t]
\begin{center}
\includegraphics[width=0.5\textwidth]{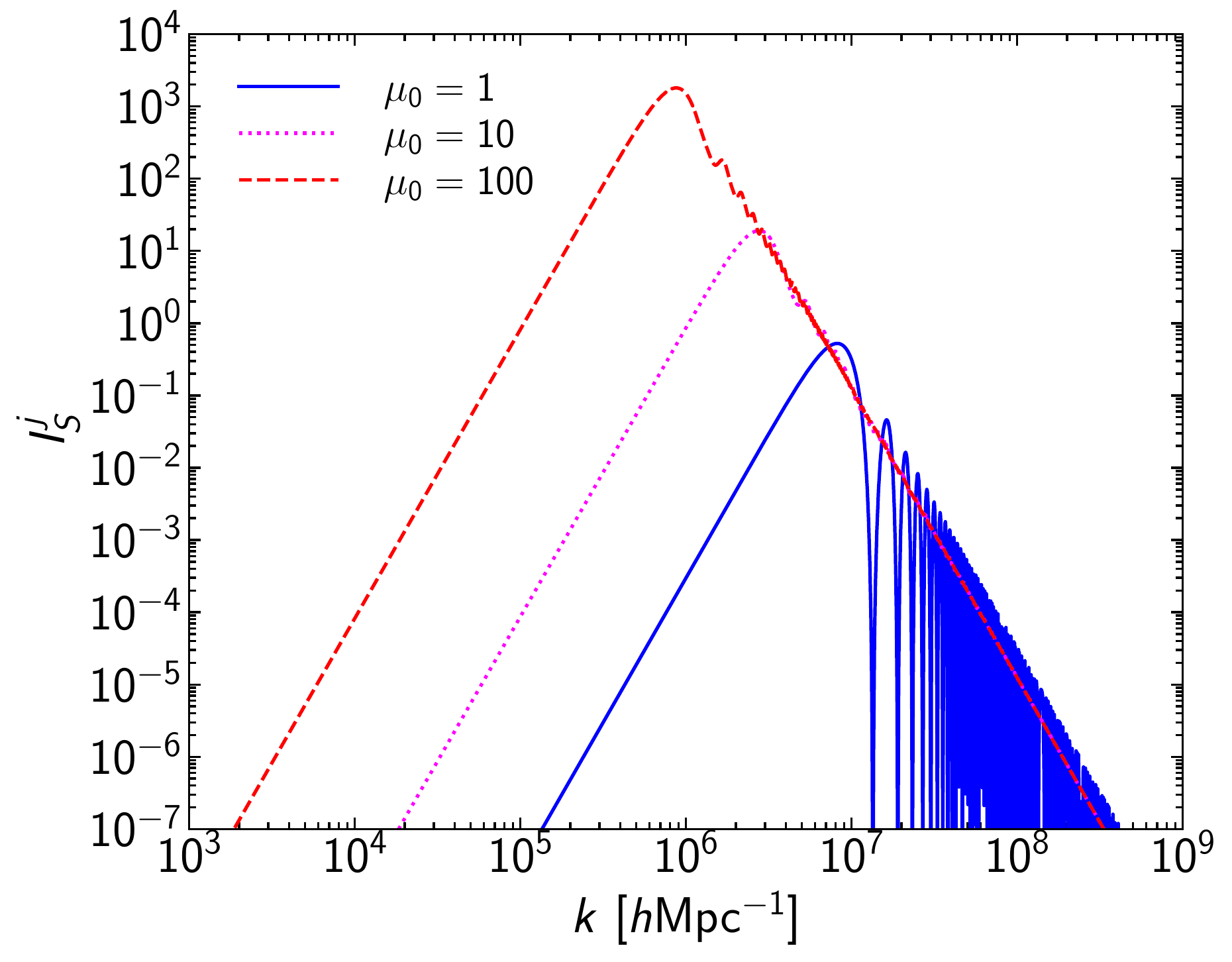}
\end{center}
\caption{The kernel function $I_S^j$ defined in Eq.~(\ref{eq:i_s}) for
  different macro model magnifications, $\mu_0=\mu_{j,1}\mu_{j,2}=1$ ({\it solid}),
  $10$ ({\it dotted}), and $100$ ({\it dashed}), which are computed changing
  $\mu_{j,1}$ while fixing $\mu_{j,2}=1$. They are computed assuming the
  lens redshift $z=0.3$, the source redshift $z_{\rm s}=2$, and the
  frequency of gravitational waves $f=10$~Hz, for which the inverse of
   the Fresnel scale is $1/r_{\rm F}\simeq 3\times 10^6 h{\rm Mpc}^{-1}$.}
\label{fig:kernel}
\end{figure}

\section{Discussions}\label{sec:discussions}

\subsection{Behavior of kernel functions}

In Sec.~\ref{sec:derivation} we show that dispersions of amplitude and
phase shifts for strongly lensed, magnified gravitational waves are
written by integrations of the convergence power spectra with kernel
functions $I_K^j$ and $I_S^j$, respectively.

As was discussed in previous work \cite{Takahashi:2005ug,Oguri:2020ldf}, 
when the macro model magnification is not considered,
$\mu_{j,1}=\mu_{j,2}=1$, kernel functions peak at around
$k\sim 1/r_{\rm F}$, which is evident from the fact that kernel
functions are described as a function of $r_{\rm F}^2k^2$. Adopting
$I_S^j$ defined in Eq.~(\ref{eq:i_s}) as a specific example, we check
how the kernel function is modified for highly magnified gravitational
waves. Fig.~\ref{fig:kernel} shows kernel functions $I_S^j$ for three
different macro model magnifications $\mu_0=\mu_{j,1}\mu_{j,2}$.
We find that the macro model magnification $\mu_0$ shifts the peak of
the kernel function such that the effective Fresnel scale becomes
larger in the presence of the macro model magnification. 
More importantly, we find that the overall amplitude of the kernel
function is significantly boosted by the macro model
magnification. The boosted amplitude and phase fluctuations can be
explained by the nonlinear dependence of convergence perturbations on
magnification factors \cite{Mao:1997ek}. We confirm that the similar
dependence on $\mu_0$ is seen for the kernel function for the
amplitude, $I_K^j$. 

Since effects of small-scale perturbations on highly magnified images
sometimes exhibit strong dependence on the image type i.e., whether the
image originates from a local minimum (Type I), a saddle point (Type
II), or a local maximum (Type III) of the arrival time surface
(e.g. \cite{Schechter:2002dm}), it is worth commenting on the possible
image type dependence of our result. We emphasize that the equations
derived in this paper do not rely on any assumption on the image type,
and hence are applicable to all image types. We check the kernel
functions for different image types to find that their dependence on
the image type is weak. For instance, when
$|\mu_{j,1}|\gg\mu_{j,2}>0$, kernel functions for positive
$\mu_{\j,1}$ and negative $\mu_{\j,1}$ are almost indistinguishable
with each other. The weak dependence on the image type is presumably
because we consider the situation that perturbations do not create
extra image pairs. In addition, our calculations are limited
to the linear order in the density perturbation, while the strong
image type dependence as mentioned above originates from the nonlinear
dependence of the magnification on convergence and shear. This means
that our calculation is valid only in the low frequency limit where
contributions from individual perturbers are significantly suppressed
due to the diffraction effect.
We also note that effects of perturbations on Type
II images may exhibit complex behaviors in the time domain as is the
case for unperturbed images (e.g., \cite{Ezquiaga:2020gdt,Dai:2017huk}).
We leave for future work more thorough studies of the dependence of the
perturbative wave optics effects on the image type.

We note that the dependence of the effective Fresnel scale and the
peak height of the kernel functions can be estimated analytically.
For the case of $I_S^j$ shown in Fig.~\ref{fig:kernel}, we find that,
when $|\mu_1|\gg|\mu_2|$, the effective Fresnel scale depends on the
macro model magnification as $\propto|\mu_0|^{1/2}$ and the peak of
the kernel function as $\propto |\mu_0|^2$. On the other hand, when
$\mu_1=\mu_2$ they exhibit different dependence on $\mu_0$ such that
the effective Fresnel scale behaves as $\propto|\mu_0|^{1/4}$
and the peak of the kernel function as $\propto |\mu_0|$.

\subsection{Dark matter substructure}
\label{sec:dm_sub}

In \cite{Oguri:2020ldf}, it was concluded that amplitude and phase
fluctuations due to dark matter substructures can be detected, if many
observations of gravitational waves from compact binary mergers are
combined statistically. Given the significant boost by the macro model
magnification $\mu_0$, amplitude and phase fluctuations may be
detected more easily for highly magnified gravitational waves, even
for a small number of events. However, as discussed in
\cite{Oguri:2020ldf}, an obstacle is microlensing by stars in lensing
galaxies or clusters of galaxies, which dominate at very small scales.

We consider a simple model of dark matter substructure to make a rough
estimate of its detectability. The small-scale power spectrum at
around the Einstein radius $r_{\rm Ein}$, where strongly lensed images
typically appear, due to dark matter substructure can be modeled as
(e.g., \cite{Hezaveh:2014aoa}) 
\begin{equation}
  P_\kappa(k)=\int dm\frac{dn}{dm}
  \left(\frac{m}{\Sigma_{\rm cr}}\right)^2\left|\tilde{y}(k)\right|^2,
\label{eq:p_kappa}
\end{equation}
where $\tilde{y}(k)$ is the Fourier transform of the normalized
Navarro-Frenk-White (NFW) density profile \cite{Navarro:1996gj},
$\Sigma_{\rm cr}$ is the critical surface density in comoving units,
and $dn/dm$ is the subhalo surface number density as a function of the 
subhalo mass $m$ at around the Einstein radius $r_{\rm Ein}$ 
\begin{equation}
  \frac{dn}{dm}=p(r_{\rm Ein})\frac{dN}{dm},
\end{equation}
with $p(r_{\rm Ein})=\Sigma_{\rm DM}(r_{\rm Ein})/M$ being the
normalized dark matter density in the projected space for the host
halo with mass $M$. We assume that the subhalo mass function $dN/dm$
follows a power-law shape
\begin{equation}
  \frac{dN}{dm}=f_{\rm
    sub}\frac{2-\alpha}{f_{\rm cut}^{2-\alpha}}
  \left(\frac{m}{M}\right)^{-\alpha}\frac{1}{M}
  \;\;\;\;\;(m\leq f_{\rm cut}M).
\end{equation}
This expression ensures that the overall mass fraction of subhalos
reduces to $f_{\rm sub}$
\begin{equation}
\int_0^{f_{\rm cut}M} m\frac{dN}{dm}dm = f_{\rm sub}M.
\end{equation} 
We adopt the slope $\alpha=1.9$, the cutoff of the subhalo mass
function $f_{\rm cut}=0.1$, and the overall mass fraction of subhalos
$f_{\rm sub}=0.1$, as typical values of these parameters
(e.g., \cite{Springel:2008cc}). 

We also follow \cite{Oguri:2020ldf} to model the effect of
stellar microlensing by the shot noise power spectrum due to the
discrete nature of stars
\begin{equation}
P_{\kappa,{\rm shot}}(k)=f_*^2\frac{M_*}{f_*\Sigma(r_{\rm
    Ein})}\left\{\kappa(r_{\rm
  Ein})\right\}^2=\frac{f_*M_*\kappa(r_{\rm Ein})}{\Sigma_{\rm cr}},
\label{eq:p_kappa_shot}
\end{equation}
where $f_*$ is the stellar mass fraction to the total mass density at
the Einstein radius, $M_*$ is the mass of individual stars (i.e., we
ignore the mass spectrum for simplicity), $\kappa(r_{\rm Ein})$ is the
convergence value (the sum of dark matter and stellar component) of
the main halo at the Einstein radius. This treatment of microlensing
is expected to be valid when the Fresnel scale is larger than the
Einstein radii of individual stars \cite{Oguri:2020ldf}.   

\begin{figure}[t]
\begin{center}
\includegraphics[width=0.5\textwidth]{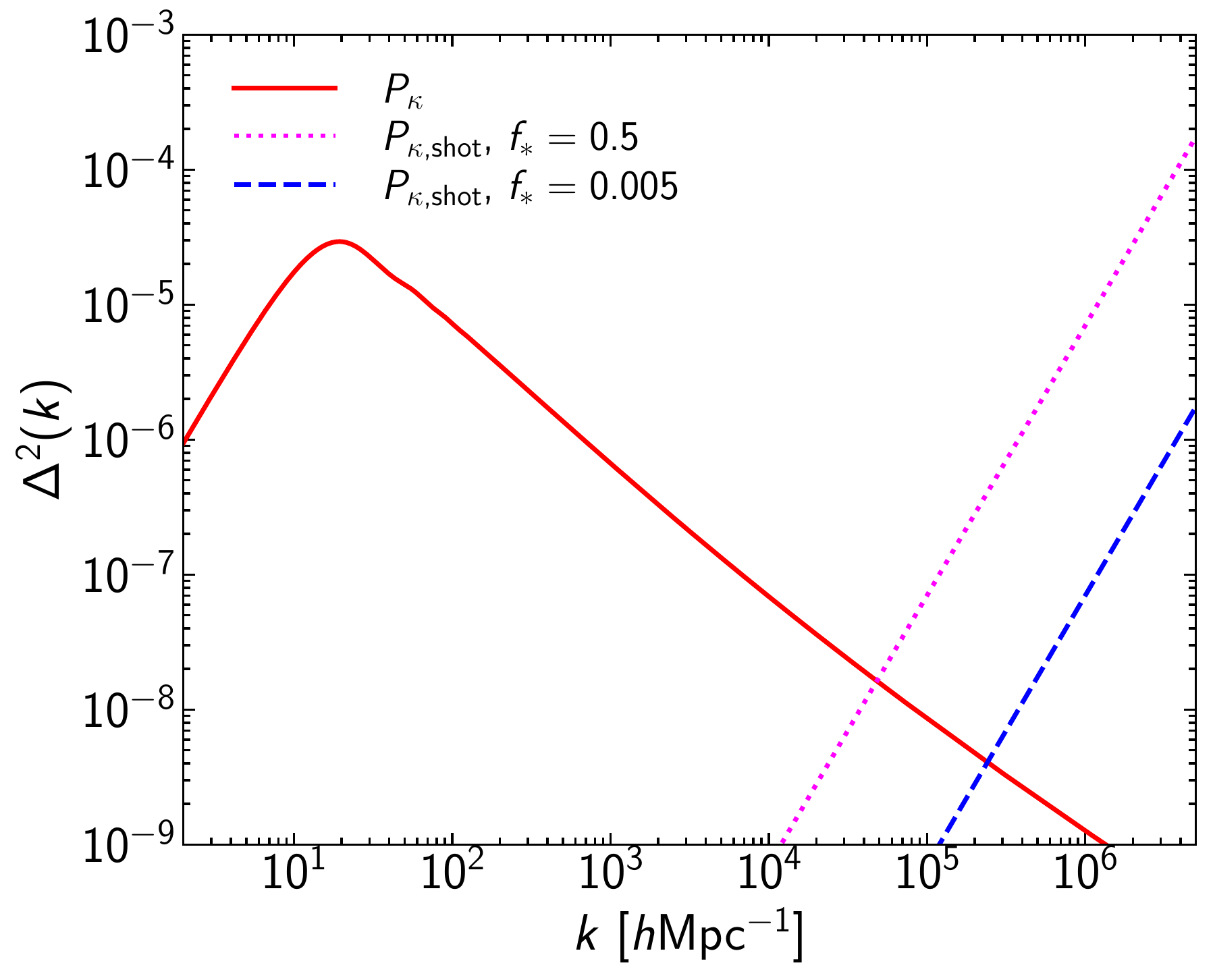}
\end{center}
\caption{The small-scale dimensionless power spectra
  $\Delta^2(k)=k^2P(k)/(2\pi)$ at around the Einstein radius due to
  dark matter substructure $P_\kappa(k)$, Eq.~(\ref{eq:p_kappa}), and
  microlensing by stars $P_{\kappa,{\rm shot}}(k)$,
  Eq.~(\ref{eq:p_kappa_shot}). See the text for parameters used for
  the calculation. For microlensing, we consider two cases with the
  stellar mass fraction of $f_*=0.5$ and $0.005$.}   
\label{fig:pk}
\end{figure}

As a specific example, we compute the convergence power spectra adopting the
host halo mass of $10^{13}h^{-1}M_\odot$ defined for the critical
overdensity of $200$, the concentration parameter of $c=10$, the lens
redshift of $z=0.3$, the source redshift of $z_{\rm s}=2$, the Einstein
radius of $r_{\rm Ein}=0.03~r_{200}$, $\kappa(r_{\rm Ein})=0.5$, and
$M_*=0.5~M_\odot$. We consider the stellar mass fraction at around the
Einstein radius of both $f_*=0.5$ and $f_*=0.005$. The former is a
typical value for galaxy-galaxy strong lens systems, while the latter
may be achieved for strong lensing due to massive clusters of galaxies
\cite{Kelly:2017fps}. Fig.~\ref{fig:pk} shows the convergence power
spectra from dark matter substructure, Eq.~(\ref{eq:p_kappa}), as well
as those from microlensing, Eq.~(\ref{eq:p_kappa_shot}).
An example of convergence power spectra shown in Fig.~\ref{fig:pk}
indicates that the effect of microlensing is dominated at the smaller
scale, or equivalently the higher wavenumber $k$. Since the macro model
magnification shifts the peak of the kernel function to the smaller
$k$, it can significantly reduce the relative contribution of
microlensing by stars to amplitude and phase fluctuations compared
with that of dark matter substructure.

\begin{figure}[t]
\begin{center}
\includegraphics[width=0.5\textwidth]{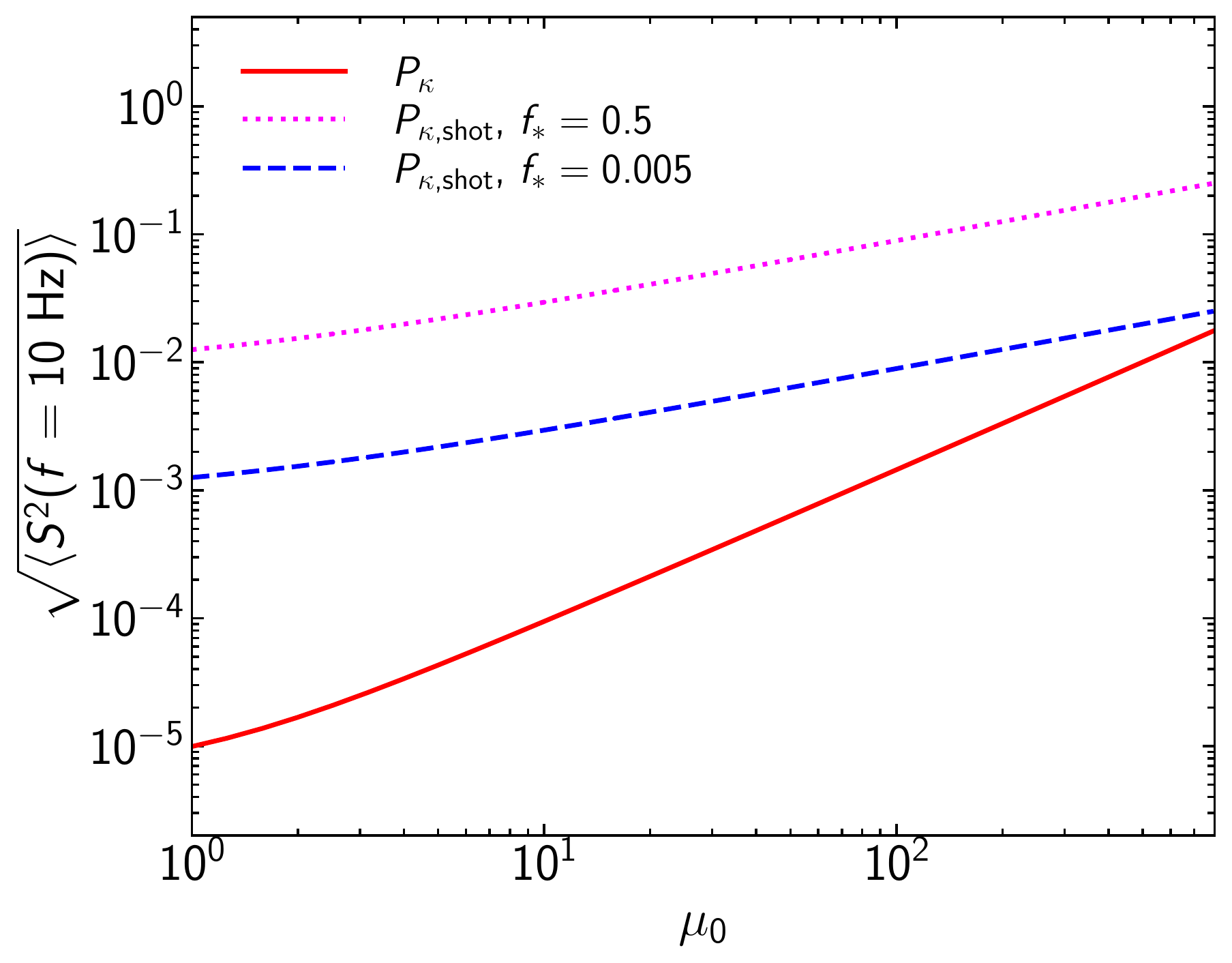}
\end{center}
\caption{Phase dispersions as a function of the macro model
  magnification $\mu_0$ computed by Eq.~(\ref{eq:s2_j}) with
  convergence power spectra of dark matter substructure
  $P_\kappa(k)$, Eq.~(\ref{eq:p_kappa}), and microlensing by stars
  $P_{\kappa,{\rm shot}}(k)$, Eq.~(\ref{eq:p_kappa_shot}), as shown in
  Fig.~\ref{fig:pk}. The frequency of gravitational waves of $f=10$~Hz
  is adopted. } 
\label{fig:var_f10}
\end{figure}

\begin{figure}[t]
\begin{center}
\includegraphics[width=0.5\textwidth]{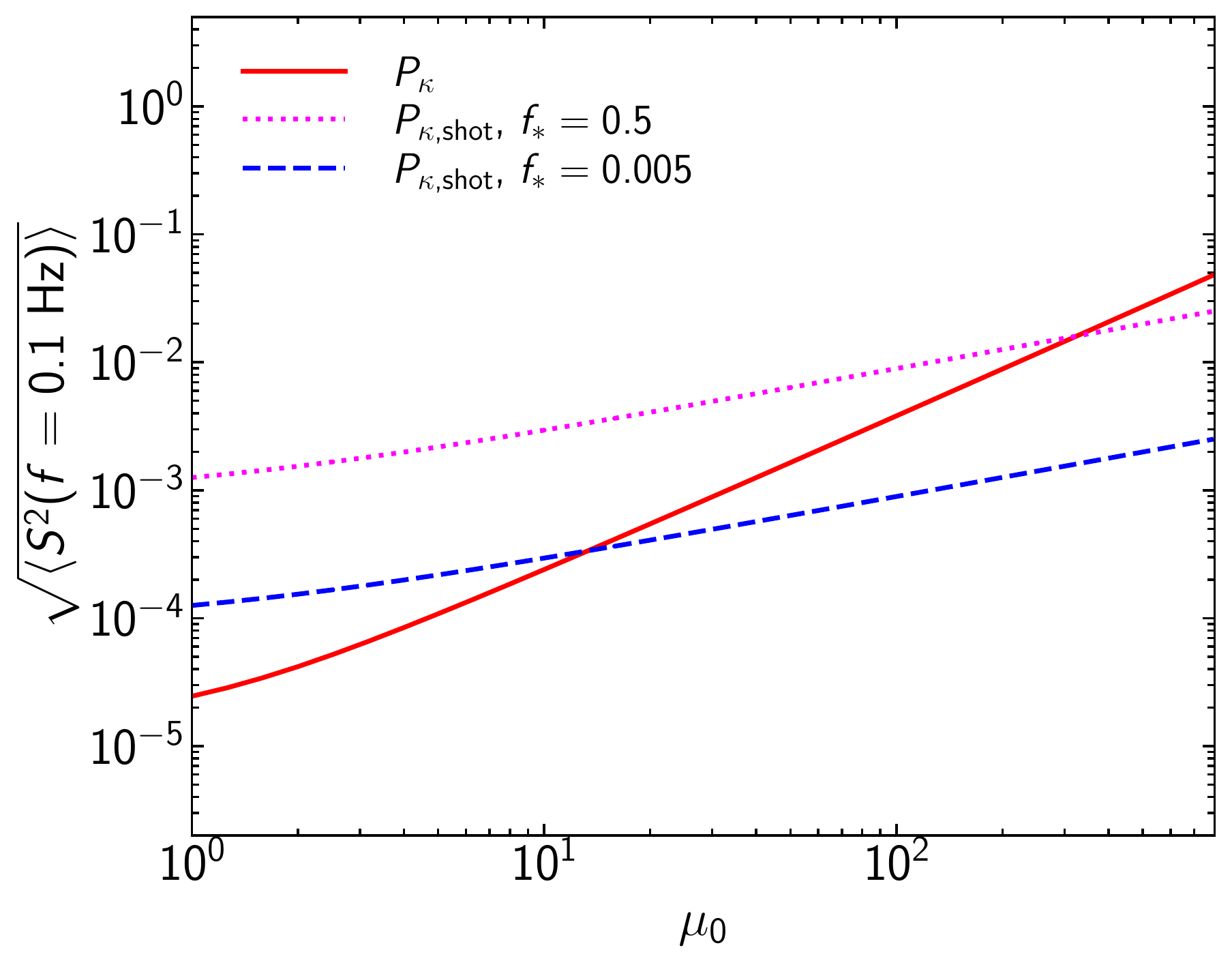}
\end{center}
\caption{Similar to Fig.~\ref{fig:var_f10}, but for $f=0.1$~Hz.}
\label{fig:var_f01}
\end{figure}

Fig.~\ref{fig:var_f10} shows phase dispersions as a function of the
macro model magnification $\mu_0$ computed by Eq.~(\ref{eq:s2_j}) for
the frequency of $f=10$~Hz that roughly corresponds to the low
frequency end of ongoing ground based laser interferometers.
In what follows, we change $\mu_0$ by changing $\mu_{j,1}$ while
fixing $\mu_{j,2}=1$ such that $\mu_0=\mu_{j,1}\mu_{j,2}=\mu_{j,1}$,
as done in Fig.~\ref{fig:kernel}.
Since the dimensionless power spectrum due to dark matter substructure
is a decreasing function of the wavenumber $k$, the phase dispersion
is quite sensitive to the macro model magnification $\mu_0$. In
contrast, the increase of the peak of the kernel function is somewhat
compensated by the shift of the peak for the case of microlensing by
stars, leading to the weaker dependence on $\mu_0$. Therefore, as
expected, the macro model magnification $\mu_0$ not only increases the
dispersion but also increases the relative contribution of dark matter
substructure to microlensing, although the contribution of dark matter
substructure appears to be subdominant even for very high $\mu_0$ at
least within the parameter range we examined.

Since the effect of microlensing is more pronounced at the higher
wavenumber $k$, it is expected that we can probe dark matter
substructure more easily at lower frequency $f$. This point is clear
from Fig.~\ref{fig:var_f01} in which results for $f=0.1$~Hz, which can
be observed by space based laser interferometers, are shown. In this
case, the contribution of dark matter substructure can become
dominant, if the macro model magnification is sufficiently large and
the stellar mass fraction is reasonably small. When the macro model
magnification is sufficiently large, several hundreds, the typical
phase shift can reach on the order of $0.1-0.01$, which can be
detected from a single event if the signal-to-noise ratio of the
observation is sufficiently high. We note that the expected
magnification of strong lensing events detected by e.g.,
advanced LIGO, advanced VIRGO, and KAGRA tends to be high due to a
strong selection effect \cite{Oguri:2018muv}, suggesting that such
high magnification events may not be rare.

Our analysis indicates that microlensing by stars can induce
non-negligible amplitude and phase shifts of highly magnified,
multiply imaged gravitational waves. Such shifts, if significant,
might degrade the performance of searching for multiple image pairs of
gravitational wave events based on the similarly of waveform shapes.
This effect might also degrade the application of using gravitational
waves to constrain modified dispersion relations
\citep{Will:1997bb,Chung:2021rcu,Ezquiaga:2022nak}.

\subsection{Fuzzy dark matter and primordial black holes}

\begin{figure}[t]
\begin{center}
\includegraphics[width=0.5\textwidth]{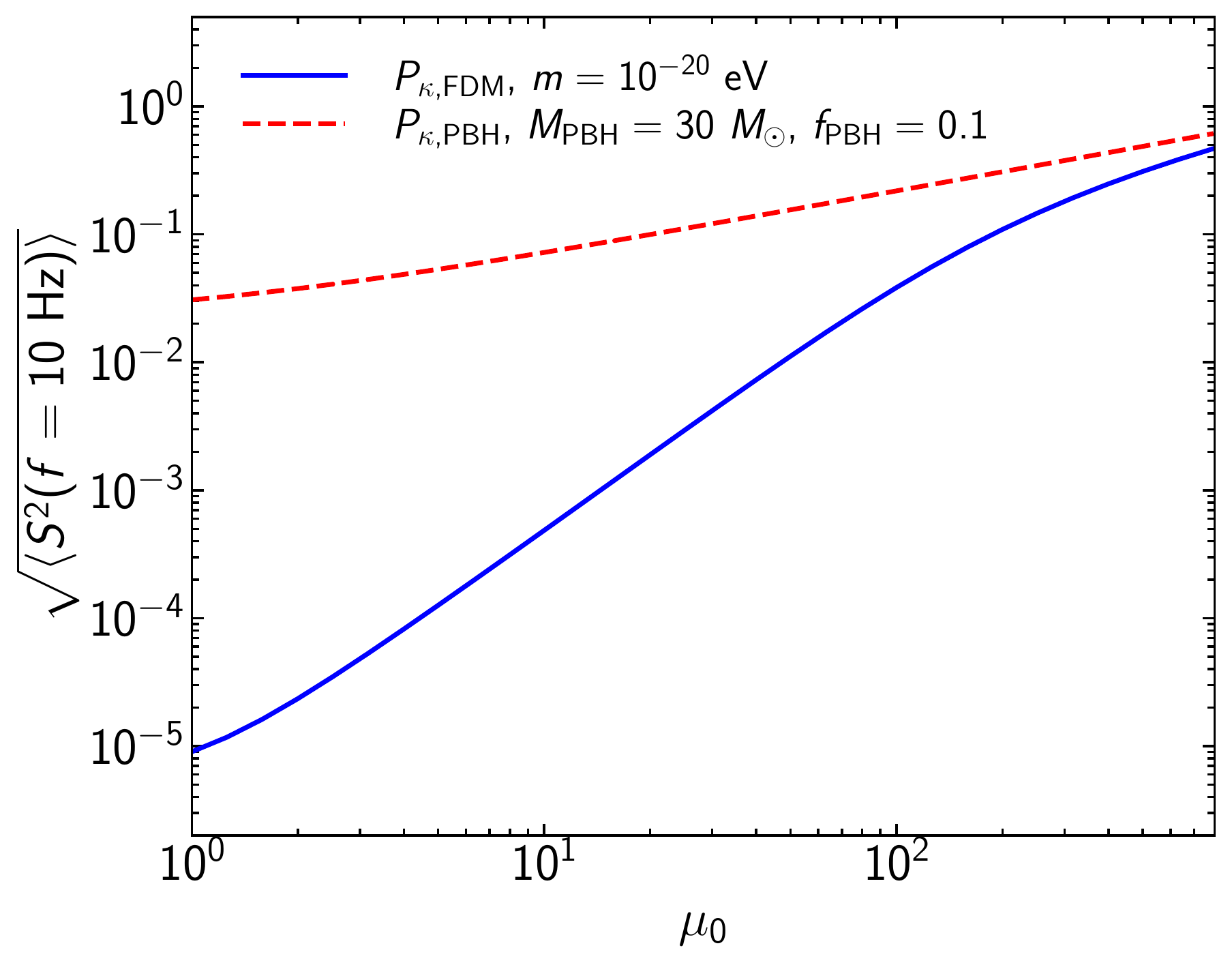}
\end{center}
\caption{Phase dispersions as a function of the macro model
  magnification $\mu_0$ in the FDM ({\it solid}) and PBH ({\it
    dashed}) models for the frequency of $f=10$~Hz.
  The calculation in the FDM model is done using
  the convergence power spectrum shown in Eq.~(\ref{eq:pk_fdm}) with
  the FDM mass of $10^{-20}$eV, while that in the PBH model is based on
 the convergence power spectrum shown in Eq.~(\ref{eq:pk_pbh}) with
 $f_{\rm PBH}=0.1$ and $M_{\rm PBH}=30~M_\odot$. Here we assume the
 stellar mass fraction of $f_*=0.5$, and the other parameters are same
 as those used in Sec.~\ref{sec:dm_sub}.}
\label{fig:var_f10_dm}
\end{figure}

The small-scale matter power spectrum can test various dark matter
models. For instance, Kawai {\it et al.} \cite{Kawai:2021znq} proposed
an analytic model of the small-small matter power spectrum in the
Fuzzy Dark Matter (FDM) model. In this model the convergence power
spectrum is given by
\begin{align}
  P_{\kappa,{\rm FDM}}(k)=&(1-f_*)^2\left\{\kappa(r_{\rm Ein})\right\}^2
  \frac{4\pi\lambda_{\rm c}^3}{3r_{\rm h}(r_{\rm Ein})}\nonumber\\
  &\times \exp\left(-\frac{\lambda_{\rm c}^2k^2}{4}\right),
  \label{eq:pk_fdm}
\end{align}
where $\lambda_{\rm c}$ is the de Broglie wavelength that is computed
from the virial velocity of the host halo and $r_{\rm h}$ denotes an
effective size of the halo defined by
\begin{equation}
  r_{\rm h}(r_{\rm Ein})=\frac{\Sigma_{\rm DM}^2(r_{\rm Ein})}{\int dz
  \,\rho_{\rm DM}^2(\sqrt{r_{\rm Ein}^2+z^2})}.
\end{equation}
Since in this paper the Fresnel scale and the wavenumber are defined
in the comoving coordinate, we compute the quantities used above also
in the comoving coordinate.

In addition, Oguri and Takahashi \cite{Oguri:2020ldf} discussed how
the small-scale matter power spectrum is modified in the Primordial
Black Hole (PBH) scenario. Here we ignore the enhancement of the halo
formation due to the isocurvature perturbation, which turns out to be
a relatively minor effect, and consider the shot noise due to the
discrete nature of PBHs
\begin{equation}
  P_{\kappa,{\rm PBH}}(k)
  =\frac{f_{\rm PBH}(1-f_*)M_{\rm PBH}\kappa(r_{\rm Ein})}{\Sigma_{\rm
      cr}},
  \label{eq:pk_pbh}
\end{equation}
where $f_{\rm PBH}$ and $M_{\rm PBH}$ denote the relative abundance of
PBHs with respect to the total dark matter density and the mass of
each PBH, respectively.

Fig.~\ref{fig:var_f10_dm} shows examples of phase fluctuations in the
FDM and PBH models to demonstrate how they can give dominant
contributions to the predicted phase fluctuations. In these examples
the phase fluctuations become sufficiently large at high $\mu_0$ such
that they dominate the contribution from stellar microlensing (see
also \cite{Diego:2019rzc} for more detailed studies on constraining
PBHs with microlensing). This result indicates that amplitude and
phase fluctuations of highly magnified gravitational waves may offer a
powerful means of constraining dark matter models. We leave more
detailed analysis on such applications to future work.

\section{Summary}\label{sec:summary}

We have derived equations that connect the small-scale density
perturbations on the Fresnel scale with amplitude and phase
fluctuations of gravitational waves in the presence of macro model
magnifications for which geometric optics is valid. We have found that
amplitude and phase fluctuations due to perturbative wave optics
effects are significantly boosted by macro model magnifications, which
not only effectively increase the Fresnel scale but also increase the
peak amplitudes of the kernel functions that are used to calculate the
amplitude and phase fluctuations. We have found that amplitude and
phase fluctuations due to dark matter substructure tend to be smaller
than those from microlensing by stars, although the former could be
detected for highly magnified gravitational waves in low frequencies,
say $f=0.1$~Hz. We have also discussed amplitude and phase
fluctuations in FDM and PBH  models, and showed that they can produce
large amplitude and phase fluctuations depending on model
parameters. While in this paper we have focused on gravitational
waves, the equations derived in this paper are not restricted to
gravitational waves and can be applied to any sources.
A caveat is that our results are based on linearized
 equations and hence ignore any nonlinear effects, although we expect
 that our formalism can be applied to a wide range of problems because
 the diffraction due to wave optics effects generally suppresses
 such nonlinear effects.

\begin{acknowledgments}
We thank the anonymous referee for useful suggestions.
This work was supported by JSPS KAKENHI Grant Numbers JP20H04725,
JP20H00181, JP20H05855, JP20H05856, JP20H04723, JP22K18720.
\end{acknowledgments}

\end{document}